\newcommand {\be}{\begin{equation}}
\newcommand {\ee}{\end{equation}}
\newcommand {\la}{\label}
\newcommand {\r}{\ref}
\newcommand {\bea}{\begin{eqnarray}}
\newcommand {\eea}{\end{eqnarray}}

\documentstyle[aps]{revtex}

\begin{document}
\title{Hole motion in the Ising antiferromagnet: an application of the
recursion method}
\author{Oleg A. Starykh\cite{a,oas} and George F. Reiter}
\address{Texas Center for Superconductivity and Physics Department,
 University of Houston, Houston, TX 77204-5932}
\date{11 July 1995}
\maketitle{}
\begin{abstract}
We study hole motion in the Ising antiferromagnet using the recursion method.
Using the retraceable path approximation we find the hole's Green's function
as well as its wavefunction for arbitrary values of $t/J_z$.
The effect of small transverse interaction also is taken into account.
Our results provide some additional insight into the
self-consistent Born approximation.

\end{abstract}

\section{Introduction}
\la{intro}
 The problem of the hole motion in a quantum antiferromagnet has
become one of the central issue in developing the theory of high-$T_c$
superconductivity. The problem is an old one \cite{BR,BNK}, but
only recently has a good understanding of it been reached (for a
review see, e.g.,\cite{yulu}).
The self-consistent Born approximation \cite{svr,KLR}
turns out to be extremely successful in predicting the energy of the
quasiparticle\cite{Man,Ho}, mainly due to
the vanishing of the low order vertex corrections \cite{Man}. Results obtained
within this approximation agree well with exact diagonalization studies
on small clusters \cite{corr-num}. The basic feature is that hole
motion is strongly renormalized by the cloud of spin excitations
(distortions) it causes, and results in a narrow (of the order of the
superexchange constant $J$) band with minima at  the $(\pm \pi/2, \pm \pi/2)$
points on the boundary of the magnetic Brilloin zone.
 More recent work by one of us \cite{reiter} has obtained the wave function
of the hole
within the same approximations and is also in agreement with exact
diagonalization results \cite{ramsak}.

 If the radius of the magnetic polaron is small enough, which happens
at $J \sim t$, its wavefunction can be constructed variationally in
direct space \cite{polaron1,polaron2}, and  the bandwidth becames
of order $t^2/J$.

 It was found long ago that hole motion in antiferromagnet with strong
Ising anisotropy produces a string of the overturned spins, which
 confines hole to the origin of its path \cite{BNK}, making the hole's
motion completely incoherent. The possibility
of curing the spin background by the hole going one and a half times around
an elementary loop on the lattice \cite{trug}
does not change this picture significantly, as the effective mass of such
coherent motion is extremely high. Surprisingly, the string picture
seems to survive the limit of the isotropic Heisenberg model, where
quantum fluctuations that can cure overturned spins are most effective.
 This shows up in the subleading peaks above the
quasiparticle peak in the hole spectral function \cite{Man,corr-num}.

 The subject of the present work is to study the hole
motion in an antiferromagnet with strong Ising anisotropy using the recursion
method. The method, also known as Lanczos technique, was developed initially
for the electronic structure calculations of disordered systems
\cite{recur,memory},
and later generalized for the finite-lattice calculations of the
strongly correlated systems \cite{corr-num,rec_ref}.
We show that the method is particularly well suited for the problem
at hand, and derive many of the known as well as some new
results within its framework.
To keep things as simple as possible we treat the $t-J$ model
within the linear spin
wave approximation, which has no formal justification in the Ising limit.
Nevertheless, we show that physics of the problem remains essentially
unchanged by this drastic approximation. Moreover, the spin wave
formalism permits us an important direct comparison with the results
of the self-consistent Born approximation approach \cite{KLR}.

 The recursion method consists in the following \cite{recur}.
Given the Hamiltonian and the initial vector $|1>$, it
generates a new basis of vectors according to the rule
\be
|n+1> = H|n> - a_n |n> - b_n^2 |n-1>,
\la{02}
\ee
with $|0>=0$, and $b_1^2 = <1|1>$. The coefficients in the recurrence are
calculated from
\be
a_n = \frac{<n|H|n>}{<n|n>}, ~~b_n^2 = \frac{<n|n>}{<n-1|n-1>}.
\la{01}
\ee
Remarkably, in the new basis the Hamiltonian has a tridiagonal form
\be
H_{nn} = a_n,~~H_{n-1,n}=H_{n,n-1}= b_n,~~H_{nm} =0 ~\text{for all}
{}~|n - m| > 1,
\la{03}
\ee
and thus describes a fictitious semi-line of "atoms" with local
orbitals $|n>$, diagonal energies $a_n$, and hopping elements
$b_n$ to the $n+1$ "atom". Then the diagonal Green's function
$G_{11}(\omega)= <1|(\omega - H)^{-1}|1>$ takes the form of a
continued fraction
\be
G_{11}(\omega)=\frac{b_1^2}{\omega - a_1 - \frac{b_2^2}{\omega - a_2
-...}}
\la{0}
\ee
We derive analytical expressions for the hole's Green's function
and wavefunction in the next Section.
The effect of small transverse interaction $J_{\perp}$ is taken into account
in Section \ref{perp}.
We find that hole motion becomes coherent, and describe it for
different limits of $t/J_z$ ratio. Comparison with previous works
is done in the Conclusion.

\section{The Ising limit}
\la{ising}
 Let us consider the motion of the hole coupled to the localized spins
on the lattice \cite{BR,BNK}. In the linear spin-wave
approximation the Hamiltonian of the
 $t-J$ model reads \cite{KLR,Man,Ho}
\bea
H= &&zt\sum_{k,q} \phi(k,q)\left(c_k^{+} c_{k-q} a_q + c_{k-q}^{+} c_k a_q^{+}
\right) + J_z\sum_k \omega(k) a_k^{+} a_k,\\&&
\phi(k,q)=u_q \gamma_{k-q} + v_q \gamma_k,\nonumber\\&&
\omega(k)=\sqrt{1 - \alpha^2 \gamma_q^2},
{}~~\gamma_q=\frac{1}{2}(\cos(q_x) + \cos(q_y))\nonumber
\eea
where  we absorbed a factor $zS$ in our definition of $J$ ($z$ is a number of
nearest neighbours),
$u_q$ and $v_q$ are the usual Bogolubov coefficients
\be
u_q=\sqrt{\frac{1 + \omega(q)}{2\omega(q)}},\\
{}~~v_q=-sgn(\gamma_q) \sqrt{\frac{1 - \omega(q)}{2\omega(q)}},
\la{A2}
\ee
 and $\alpha\equiv J_{\perp}/J_z$.
Each hopping of the hole produces emission or absorption of the spin
excitations.

For the problem of single hole motion in antiferromagnet
it is natural to choose as a starting
vector the state with one hole and no spin deviations,
i.e. $|1> = c_p^{+}|vac>$,
where $|vac>$ denotes vacuum for both hole and magnon operators.
Thus $|vac>$ is simply the Neel state.
 The next vector is then
\be
|2> = H|1> = H_t|1> = zt\sum_q \phi(p,q) c_{p-q}^{+} a_q^{+}|vac>
\la{A3}
\ee
Clearly, $<1|2> =0$, and $b_2^2 = \frac{<2|2>}{<1|1>} = (zt)^2 \sum_q \phi^2
(p,q)$.
 Acting by $H$ on $|2>$ one finds
\bea
&&H_t|2> =(zt)^2\sum_q \phi^2(p,q) c_p^{+}|vac> + (zt)^2 \sum_{q,q_1} \phi(p,q)
\phi(p-q,q_1) c_{p-q-q_1}^{+} a_{q_1}^{+} a_q^{+} |vac>\nonumber\\&&
H_J|2> = ztJ\sum_q \omega(q) \phi(p,q) c_{p-q}^{+} a_q^{+}|vac>\\&&
a_2= J \frac{\sum_q \omega(q) \phi^2(p,q)}{\sum_q \phi^2(p,q)}\nonumber
\eea
Following the recipe described in the Introduction one obtains
\bea
|3> =&& (zt)^2 \sum_{q,q_1} \phi(p,q)\phi(p-q,q_1) c_{p-q-q_1}^{+}
a_{q_1}^{+} a_q^{+} |vac>\nonumber\\&&
+ ztJ\sum_q \left(\omega(q) - \frac{\sum_q \omega(q) \phi^2(p,q)}
{\sum_q \phi^2(p,q)}\right) \phi(p,q) c_{p-q}^{+} a_q^{+}|vac>
\la{A5}
\eea
 The procedure becomes untractable very fast because of the branching:
each action of $H$ on a state with $n$-magnons produces again a state with
$n$-magnons ($H_J$ term) as well as states with $(n+1)$- and $(n-1)$-magnons
($H_t$ term), whose coefficients have a complicated momentum dependence.
The exception occurs when $\omega(q)$ is momentum-independent, i.e. in the
Ising ($\alpha=0$) case. In that case one has $a_2 =J_z$ and
\be
|3> = (zt)^2 \sum_{q,q_1} \phi(p,q)\phi(p-q,q_1) c_{p-q-q_1}^{+}
a_{q_1}^{+} a_q^{+} |vac>,
\la{A6}
\ee
where now $\phi(p,q)=\gamma_{p-q}$, and $u_q=1,~~v_q=0$.
 The norm of this state is
\bea
<3|3> =&& (zt)^4 \sum_{q,q_1} \sum_{l,l_1} \phi(p,q) \phi(p-q,q_1) \phi(p,l)
\phi(p-l,l_1)<vac|a_l a_{l_1} a_{q_1}^{+} a_q^{+} |vac>\nonumber\\
&&=(zt)^4 \sum_{q,q_1} \phi^2(p,q) \phi^2(p-q,q_1)
\la{A7}
\eea
due to the property $\sum_{q_1} \gamma_{p-q_1} \gamma^2_{p-q-q_1} =0$.
Thus, among two possible pairings of $a$-operators in (\ref{A7}) only
the "diagonal" one $(l_1 =q_1,~l=q)$ contributes. Note also that
$\sum_{q_1} \gamma^2_{p-q-q_1} =\frac{1}{z}$,
and hence $b_3^2 = zt^2$.
 Continuing one finds
\bea
&&a_3=J_z \frac{\sum_{q,q_1}\left(\omega(q) + \omega(q_1)\right)
\phi^2(p,q)\phi^2(p-q,q_1)}{\sum_{q,q_1} \phi^2(p,q) \phi^2(p-q,q_1)}
=2J_z~,\nonumber\\
&&|4>= (zt)^3\sum_{q,q_1,q_2} \phi(p-q-q_1,q_2)\phi(p-q,q_1)\phi(p,q)
c^{+}_{p-q-q_1-q_2} a^{+}_{q_2} a^{+}_{q_1} a^{+}_q |vac>
\la{8}
\eea
 Now, however, in addition to the "diagonal" pairing $(l_i = q_i)$
 a new one
$(l=q_2,~l_1=q_1,~l_2=q)$ appears, for example,
\be
<4|4>=(zt)^6\left\{ (\frac{1}{z})^3 + \sum_{q,q_1,q_2} \gamma_{p-q}
\gamma_{p-q_2} \gamma_{p-q_1 -q_2} \gamma_{p-q-q_1}\gamma^2_{p-q-q_1 -q_2}
\right\}
\la{44}
\ee

 This pairing describes hole jumping around elementary loop
(plaquette on a square lattice) 1.5 times \cite{trug}.
 The triple integral in (\ref{44}) is calculated to give
$(\frac{1}{z})^3 ~\frac{1}{16}~\left(2 + \cos(p_x - p_y) \right) \left(2 +
\cos(p_x + p_y) \right)$. It describes an effective hole hopping along the
same sublattice. The hopping is maximal for $p=(0,0)$, and minimal for
$p=(\pi,0)$ or $(0,\pi)$. Propagation via closed loops was studied in
great detail in Ref.\cite{geometry}. For $J_z \gg t$ the effective
hopping corresponds to the tunneling through a potential barrier,
(it scales as $\exp\left[-4.6 (J_z/t)^{1/2}\right]$ \cite{geometry}),
and leads to the band with
minima at $p=(0,0)$. In the opposite limit $(t \gg J_z)$ hole can
travel very far before noticing the confining potential due to the
finite $J_z$, and the weight of the closed-loop paths among the long
self-retraceable paths is very small. This point of view is supported
by the results of \cite{ss}, where it was found that the ground state energy
of the one-hole system scales as $J_z^{2/3}$ in a wide range of
$J_z/t$ $(1> J_z/t > 5 \cdot 10^{-3})$, which is a characteristic
feature of the retraceable-path approximation (see our discussion below).

Based on these arguments we will omit all closed-loop contribution in
the following. This corresponds
to solving the problem on a Bethe lattice, or in the limit of infinite
dimensionality \cite{vollhardt}.  Diagrammatically this approximation
consists in neglecting the vertex corrections \cite{Man,Ho}.

With this simplification in mind the procedure can be iterated
infinitely and results in the following expression for the coefficients
of the recursion
\be
a_n = (n-1)J_z,~~b_n^2 = zt^2,~~{\text for~any~} n \geq 2.
\la{9}
\ee
 As should be clear by now, linear $n$-dependence of $a_n$ comes
from the simple fact that $H_J |n>= (n-1) J_z |n>$.
On the other hand the $n$-independence of $b_n$ is due
to the particular form of the coupling $\gamma_{p-q}$. The strong momentum
dependence of this coupling favors only one particular pairing
out of $(n-1)!$ possible at the $n$-th step of the recursion procedure.
 This particular pairing is nothing
but the retraceable path approximation of Brinkman and Rice \cite{BR}:
first the
last excited magnon is absorbed, then the one before last, etc...
If the hole-boson coupling is momentum independent, as happens, for example,
in the Holstein model, the outlined procedure leads to $b_n^2 \sim n$
\cite{holstein}, and, as a result, to a different physics.

 So far we have found that Hamiltonian $H$ has a simple tridiagonal
form in the
space spanned by the vectors $|n>$ (see Eq.(\ref{03})).
Then for any vector $|\Psi>=\sum_n \Psi_n |n>$ of this
space the Schrodinger equation $H|\Psi>= E|\Psi>$ holds, or
\be
(E - a_n)\Psi_n - b_{n-1}\Psi_{n-1} - b_{n+1} \Psi_{n+1} =0,
\la{10}
\ee
with boundary condition $\Psi_0 =0$ (remember that $|0>=0$ by construction).
In small-$J$ (large-$n$) limit we may use a continuum approximation to find
\be
\pm \sqrt{z}t \frac{\partial^2 \Psi_n}{\partial n^2} \pm 2\sqrt{z}t\Psi_n
+(n-1)J_z \Psi_n = E\Psi_n
\la{11}
\ee
Which is just the Schrodinger equation for a particle on a semi-line
in the linear potential. Interestingly, index $n$ plays a role of
coordinate along the path of the hole - this correspondence is exact
because we neglected all closed loops. The $+(-)$ sign
in (\ref{11})
corresponds to choosing energy E close to the upper (lower) edge of the band.
At the bottom of the band one finds $J_z^{2/3}$-behaviour:
\be
E_n = -2\sqrt{z}t - J_z + \beta_n J_z^{2/3}(\sqrt{z}t)^{1/3}
\la{12}
\ee
here $-\beta_n$ are the zeros of Airy function.
It is known that the spin-wave approximation overestimates the bandwidth
\cite{KLR}, and more accurate treatment of the constraint leads
to the replacement of $\sqrt{z}$ by the $\sqrt{z-1}$ in the
formulas above.

The result obtained is by no means new and goes back to the sixties \cite{BNK}.

We would like to demonstrate a different approach here.
Consider the one-hole Green's function, which has a continued
fraction form (see Eq.(\ref{0})).
\be
G_{11}(\omega)=\frac{1}{\omega - \frac{zt^2}{\omega - J_z -\frac{zt^2}
{\omega - 2J_z-...}}}
\la{13}
\ee
A little thinking shows that it is equivalent to
\be
G(\omega)(\omega -zt^2 G(\omega - J_z))=1
\la{14}
\ee
which coincides with the self-consistent Born approximation
 \cite{KLR}, in agreement with our expectations. Let us
try the ansatz \cite{foget}
\be
G(\omega)=-\frac{1}{\sqrt{z}t}\frac{\Upsilon(\omega)}{\Upsilon(\omega + J_z)}.
\la{15}
\ee
Then Eq.(\r{14}) takes the form of the difference equation
\be
\Upsilon(\omega - J_z) + \Upsilon(\omega + J_z) = -\frac{\omega}{\sqrt{z}t}
\Upsilon(\omega),
\la{16}
\ee
 which is a well-known recursion relation for the Bessel
functions \cite{gradst}, and hence
$\Upsilon(\omega)=A
 J_{-\frac{\omega}{J_z}}\left(\frac{2\sqrt{z}t}{J_z}\right)$.
This immediately gives for the Green's function
\be
G_{11}(\omega)=-\frac{1}{\sqrt{z}t}
\frac{J_{-\frac{\omega}{J_z}}\left(\frac{2\sqrt{z}t}{J_z}\right)}
{J_{-\frac{\omega + J_z}{J_z}}\left(\frac{2\sqrt{z}t}{J_z}\right)}
\la{18}
\ee
This form is a direct consequence of the linear dependence of $a_n$
on $n$ (see Ref.\cite{matan}), which, in turn, is an intrinsic
feature of the retraceable path approximation.
 A similar expression was obtained within the retraceable-path approximation
(without using spin-wave transformation) in Ref.\cite{mohan}.
Notice that energy appears only in the indexes of the Bessel functions.

Let us look on the $\frac{t}{J_z} \gg 1$ limit first.
Here we need the double asymptotic expansion of the Bessel function
at large value of argument and index \cite{gradst}
\bea
&&J_{\mu \cos\theta}(\mu) \sim \sqrt{\frac{2}{\pi\mu\sin\theta}}
\cos\left\{\mu (\sin\theta - \theta\cos\theta) -\frac{\pi}{4}\right\}
\\&&
\mu > 0,~~0 < \theta < \pi/2 \nonumber
\eea
Somewhat lengthly calculations give an expected answer
\bea
&&G_{11}(\omega)=\frac{J_z}{\sqrt{z}t} \sum_{\nu=0}^{\infty}
{}~\frac{1}{\omega - \omega_{\nu}}\\&&
\omega_{\nu} = -2\sqrt{z}t - J_z +
\beta_{\nu} J_z^{2/3}(\sqrt{z}t)^{1/3}, \nonumber
\eea
in complete agreement with results of the Schrodinger equation (\ref{12}).
Notice that residue of the poles is $\nu$-independent, and is given
by $J_z\over{\sqrt{z}t}$, supporting the "dominant-pole" approximation
of \cite{KLR}. We note in passing that the quasiclassical
expression for the zeros of the Airy function
$\beta_{\nu}=\left(\frac{3\pi}{2}(\nu + \frac{3}{4})\right)^{2/3}$
works extremely well even for $\nu=0$: it gives $\sim 2.32$ compared to
the exact value $\beta_0 = 2.34$.

In the opposite $\frac{t}{J_z} \ll 1$ limit one can use standard
small argument expansion
\be
J_{\eta}(x)=\left(\frac{x}{2}\right)^{\eta} \sum_{k=0}^{\infty}
\frac{(-1)^k}{k!\Gamma(\eta + k + 1)} \left(\frac{x}{2}\right)^{2k},
\ee
 to find the perturbation theory result
$G_{11}(\omega)=(\omega + \frac{zt^2}{J_z})^{-1}$.

Recently one of us \cite{reiter} derived a wavefunction that
corresponds exactly to the self-consistent Born approximation,
\bea
|\Psi_k^{\text{scba}}> = && z_k \big(c_k^{+} + N^{-1/2}\sum_q zt \phi(k,q)
G_{k-q}(\epsilon_k -\omega_q) c_{k-q}^{+} a_q^{+} +... \\&&
+N^{-n/2}\sum_{q,q_1,...q_{n-1}} (zt)^{n}
\phi(k,q)G_{k-q}(\epsilon_k -\omega_q)...
\phi(k-q-q_1 -...-q_{n-2},q_{n-1})\nonumber\\&&\times
G_{k-q-q_1 -...-q_{n-1}}(\epsilon_k -\omega_q
-...-\omega_{q_{n-1}}) c_{k-q-q_1 -...-q_{n-1}}^{+} a_q^{+}...a_{q_{n-1}}^{+}
\big)|vac>.\nonumber
\eea
Here $\epsilon_k=\Sigma_k(\epsilon_k)$ is the quasiparticle energy, and
$\Sigma_k$ is the self-energy in the SCBA. The quasiparticle spectral weight
is given by $(z_k)^2 = \left(1 - \frac{\partial \Sigma_k(\omega)}{\partial
\omega} |_{\omega=\epsilon_k}\right)^{-1}$.
 Clearly, in our case the role of the momentum $k$ is played
by the index $\nu$.
The specific form of the Green's function (\ref{18}) immediately gives us
that the chain product of the Green's function in the last equation
has a simple form
$G(\omega - J_z)G(\omega - 2J_z)...G(\omega - nJ_z)= (\frac{-1}{\sqrt{z}t})^{n}
J_{-\frac{\omega - nJ_z}{J_z}}(\frac{2\sqrt{z}t}{J_z})/J_{-\frac{\omega}{J_z}}
(\frac{2\sqrt{z}t}{J_z})$.
Using again asymptotic expansion of the Bessel functions
we find for the hole in lowest state $\nu=0$ ($t/J_z \gg 1$)
\bea
|\Psi_k^{\text{scba}}> \sim &&\sqrt{\frac{J_z}{\sqrt{z}t}}\big(c_k^{+} -
N^{-1/2}\sum_q \sqrt{z} \gamma_{k-q}~c_{k-q}^{+} a_q^{+} + \nonumber\\
&&...+N^{-n/2}\sum_{q,q_1,...q_{n-1}} (-1)^{n} z^{n/2}~n~\gamma_{k-q}...
\nonumber\\&&\times
\gamma_{k-q-...-q_{n-1}} ~c_{k-q-q_1 -...-q_{n-1}}^{+}
a_q^{+}...a_{q_{n-1}}^{+}
\big)|vac>,
\eea
for $n \ll \left(\sqrt{z}t/J_z \right)^{1/3}$.

\section{Small $J_{\perp}$ limit}
\la{perp}

A virtue of the present approach is that we can
include the effect of small transverse interaction $J_{\perp}$,
$J_{\perp}=\alpha J_z \ll J_z$.
As was already discussed by \cite{KLR}, analytical consideration is possible
if one restricts to linear in $\alpha$ accuracy.
 Then $\omega(q)= 1 + O(\alpha^2), u_q = 1 + O(\alpha^2)$,
 but $v_q= -\frac{1}{2} \alpha \gamma_q$.
Hence
\be
\phi(p,q)= \gamma_{p-q} - \frac{1}{2} \alpha\gamma_q \gamma_p + O(\alpha^2)
\ee
One finds for first $a$- and $b$-coefficients
\be
a_1 =0,~~a_2=J_z,~~b_2^2= (zt)^2 \sum_q\left(\gamma_{p-q}^2 - \alpha
\gamma_p \gamma_q \gamma_{p-q} \right)=zt^2(1 - \alpha \gamma_p^2)
\ee
The key property $\sum_q \gamma_{p-q} \gamma_{k-q}^2 =0$ insures
that the retraceable path approximation is exact at low order, i.e. that
\be
\sum_{q,q_1} \phi(p,q_1)\phi(p-q_1,q)\phi(p-q,q_1)\phi(p,q) = 0 + O(\alpha^2).
\ee
 Moreover, one can find
\bea
I_n \equiv &&\sum_{q,q_1,...q_{n-1}} \phi^2(p,q)\phi^2(p-q,q_1)...
\phi^2(p-q-q_1 -...-q_{n-2},q_{n-1})\nonumber\\&&
=\sum_{q,q_1,...q_{n-2}} \phi^2(p,q)\phi^2(p-q,q_1)...
\phi^2(p-q-q_1 -...-q_{n-3},q_{n-2})\nonumber\\&&\times
\sum_{q_{n-1}}
\phi^2(p-q-q_1 -...-q_{n-2},q_{n-1})\nonumber\\&&
=\frac{1}{z} I_{n-1} -\frac{\alpha}{z} \left(\frac{1}{z}\right)^{n-2}
\sum_q \gamma^4_q
\eea
As a result
\bea
&&b_n^2= zt^2 (1 - \alpha z B) + O(\alpha^2),~n\geq 3\\&&
B=\sum_q \gamma^4_q = \frac{9}{64}\nonumber
\eea
At the same time there are no changes in $a_n$, $a_n = (n-1)J_z$.
Due to the difference between $b_2^2$ and $b_{n\neq 2}^2$, the
continued fraction expression for the Green's function is
now equivalent to the following system of two equations:
\bea
&&G_{11}(\omega)=\frac{1}{\omega - zt^2(1 - \alpha \gamma_p^2)G_{22}(\omega)}
\nonumber\\&&
G_{22}(\omega)=\frac{1}{\omega - J_z - z\tilde{t}^2 G_{22}(\omega - J_z)}
\la{p1}
\eea
We introduced $\tilde{t}^2 = t^2 (1 - \alpha z B)$ here.
Comparing the last equation
of (\r{p1}) with Eqs.(\r{14},\r{18}) of the previous Section
one finds the answer
\be
G_{22}(\omega)= -\frac{1}{\sqrt{z}\tilde{t}}
\frac{J_{-\frac{\omega - J_z}{J_z}}\left(\frac{2\sqrt{z}\tilde{t}}{J_z}\right)}
{J_{-\frac{\omega}{J_z}}\left(\frac{2\sqrt{z}\tilde{t}}{J_z}\right)}
\la{p2}
\ee
 In the $J_z \ll \tilde{t}$ limit expression (\r{p2}) takes familiar
form
\bea
&&G_{22}(\omega)=\frac{J_z}{\sqrt{z}\tilde{t}} \sum_{\nu=0}^{\infty}
{}~\frac{1}{\omega - \bar{\omega}_{\nu}}\\&&
\bar{\omega}_{\nu} = -2\sqrt{z}t +
\beta_{\nu} J_z^{2/3}(\sqrt{z}\tilde{t})^{1/3} \nonumber
\eea
We want to consider the lowest hole state, and thus may approximate
\bea
&& \sum_{\nu=0}^{\infty}~\frac{1}{\omega - \bar{\omega}_{\nu}} \simeq
\frac{1}{\omega - \bar{\omega}_{0}} + \sum_{\nu \neq 0}
{}~\frac{1}{\bar{\omega}_{0} - \bar{\omega}_{\nu}}\nonumber\\&&
=\frac{1}{\omega - \bar{\omega}_{0}} + \left(\frac{3\pi J_z}{2}\right)^{-2/3}
\left(\sqrt{z}\tilde{t}\right)^{-1/3}~\sum_{\nu \neq 0}
\frac{1}{\nu^{1/3}(\nu + 3/2)^{1/3}}
\eea
The last sum can be approximated by the integral from $0$ to some
$\nu_m$, which, in turn, is determined by the condition
 $\bar{\omega}_{\nu_m} \simeq 2\sqrt{z} \tilde{t}$ (upper edge of the
incoherent band at $J_z =0$).
In this way we find that last sum is given by  $- \frac{4}{\pi J_z}$.
 We substitute the result just found for $G_{22}(\omega)$
into the first of the equations (\r{p1}), and
find after some algebra
\bea
&&G_{11}(\omega) = \frac{\pi^2}{4(\pi - 2)}~\frac{J_z}{\sqrt{z}\tilde{t}}
{}~\frac{1}{\omega - \tilde{\omega}}\nonumber\\&&
\tilde{\omega} = -2\sqrt{z}\tilde{t} + 2.34 J_z^{2/3}(\sqrt{z}\tilde{t})^{1/3}
- \frac{\pi J_z}{2(\pi-2)}(1 - \alpha \gamma_p^2)
\la{p4}
\eea
The pole at $\tilde{\omega}$ describes a narrow, coherent band of
bandwidth $\sim 1.38 J_{\perp}$ with minima at $\gamma_p =0$, i.e. along
the boundary of the magnetic Brilloin zone \cite{KLR}. Note that residue
of the pole remains unaffected.
 It is a matter of short calculation to find out that in the opposite
limit ($J_z \gg \tilde{t}$) the answer is
\be
G_{11}(\omega)=\frac{1}{\omega + \frac{z\tilde{t}^2}{J_z}\left(
1 - \alpha \gamma^2_p\right)}
\ee
 The bandwidth is much smaller, $z\tilde{t}^2J_{\perp}/J_z^2$,
and goes to the correct perturbative answer $\sim zt^2/J$ at the
isotropic point $J_{\perp}=J_z$ ($\alpha=1$).

\section{Conclusions}
A simple recursion method was used to calculate a single
particle Green's function for the hole moving in the antiferromagnet
with strong Ising anisotropy. By neglecting closed-loop contribution
we reduced the problem to an exactly solvable one.
In the Ising limit we find, in agreement with known results, that
 hole is confined to the origin of its
path by the effective linear potential due to the overturned spins.
We then take into account the small transverse interaction between
spins and find that the hole motion becames coherent. We calculate
the hole's spectrum and Green's function to the first order in
$J_{\perp}/J_z$. This calculation explains the "dominant-pole" approximation
of \cite{KLR}. We argue that the spin-wave approximation, employed
in this paper, does not affect the essence of the problem, as
can be seen, for example, from the comparison
of our expression (\r{18}) for the hole's Green's function
with formula (4) of Ref.\cite{mohan}. Overall, our approach can
be considered as an another way to arrive at the
self-consistent Born approximation \cite{KLR}, at least
in the case of strong Ising anisotropy.
 Our results become exact in the limit of infinite dimension.

O.A.S. acknowledges an interesting discussion with
R.O.Kuzian, which initiated this study. We are grateful to A.L.Chernyshev
for valuable discussions and comments, and to V.Brazhnikov for help.

\end{document}